\documentstyle[multicol,aps,prb,psfig]{revtex}
\begin{document}

\title {Numerical study of the strongly screened vortex glass
  model in an external field}

\author{Frank Pfeiffer$^{1}$ and Heiko Rieger$^{1,2}$}

\address{
$^{1}$ Institut f\"ur Theoretische Physik, Universit\"at zu K\"oln, 
       50937 K\"oln, Germany\\
$^{2}$ NIC c/o Forschungszentrum J\"ulich, 52425 J\"ulich, Germany
}

\date{February 21, 1999}

\maketitle

\begin{abstract}
  The vortex glass model for a disordered high-$T_c$ superconductor in
  an external magnetic field is studied in the strong screening limit.
  With exact ground state (i.e.\ $T=0$) calculations we show that 1)
  the ground state of the vortex configuration varies {\it
    drastically} with infinitesimal variations of the strength of the
  external field, 2) the minimum energy of global excitation loops of
  length scale $L$ do {\it not} depend on the strength of the external
  field, however 3) the excitation loops themself depend sensibly
  on the field. From 2) we infer the absence of a true superconducting
  state at any finite temperature independent of the external field.
\end{abstract}


\begin{multicols}{2}
\narrowtext

\newcommand{\bc}{\begin{center}}
\newcommand{\ec}{\end{center}}
\newcommand{\be}{\begin{equation}}
\newcommand{\ee}{\end{equation}}
\newcommand{\beqn}{\begin{eqnarray}}
\newcommand{\eeqn}{\end{eqnarray}}
\newcommand{\ba}{\begin{array}}
\newcommand{\ea}{\end{array}}

\section{Introduction}

The gauge or vortex glass model has become a paradigm in studying
amorphous high-T$_c$ superconductors or random Josephson-junction
arrays (see \cite{blatter} for a review). One essential feature of this
model is the possible appearance of a glassy state at low enough
temperatures, without which true superconductivity (i.e.\ vanishing
resistance) would cease to exist in these disordered materials
\cite{Fisher,FFH}.

Experimental evidence of such a vortex glass state has been reported
for high-T$_c$ superconductors \cite{exp}. From the theoretical side
it is now commonly believed that in the absence of screening a true
superconducting vortex glass phase occurs at low enough temperatures
\cite{FTY,kost1,kost2}. If screening is present the original,
unscreened $1/r$-interaction of the vortex lines is exponentially
shielded beyond a particular length scale $\lambda$ and the situation
seems to change, in particular in the limit in which the screening
length is zero (i.e.\ where vortex lines interact only on-site) the
low temperature vortex glass phase seems to be destroyed
\cite{BY,WY,KR,kost2}.

In a typical experimental situation \cite{exp} the amorphous
high-T$_c$ superconductor is put into a homogeneous magnetic field
pointing, say, in the $z$-direction. Due to bulk disorder, i.e.\ 
inhomogeneities (vacancies, defects, etc.), in the bulk of the sample
the vector potential acting on the superconducting phase variables
attains a random component (most plausible this mechanism is explained
in the context of granular superconductors \cite{gauge}), however,
still there should be a homogeneous back ground field superposed on
the random part. 

Therefore in this paper we study the question of how is the latter
scenario, i.e.\ the absence of a true superconducting phase in the
strongly screened three-dimensional gauge or vortex glass model
influenced by the presence of a homogeneous external field in one
particular space direction. This is done via the investigation of
exact ground states of the vortex glass Hamiltonian and its low energy
excitation. First we analyze the sensibility of the minimum energy
configuration with respect to the addition of a homogeneous external
field, then we study the low energy excitations of length scale $L$ in
the spirit of the usual domain wall renormalization group (DWRG)
calculations \cite{kost1,kost2,gremp}.

The lattice model describing the phase fluctuations (described by
phase variable $\phi_i\in[0,2\pi[$) in a strongly disordered
superconductor close to a normal-to-superconductor phase transition is
the gauge glass model \cite{blatter,BY} 
\beqn
  H & = & - J \sum\limits_{\langle ij\rangle} \cos(\phi_i - \phi_j -
  A_{ij} - \lambda^{-1} a_{ij}) \cr 
  & & + \frac{1}{2}\sum\limits_{\Box} (\nabla \times {\bf a} )^2 ,
  \label{gauge}
\eeqn
where the first sum runs over all nearest-neighbor pairs $\langle
ij\rangle$ on a $L\times L\times L$ simple cubic lattice and the
second over every elementary plaquette $\Box$, respectively, and
$a_{ij}$ the fluctuating vector potentials and $\lambda$ the screening
length. $A_{ij}=A_{ij}^{\rm rand}+A_{ij}^{\rm hom}$ are the quenched
vector potentials consisting of a random component $A_{ij}^{\rm
  rand}\in[0,2\pi[$ and a homogeneous component $A_{ij}^{\rm hom}$
modeling an external magnetic field in the $z$-direction.  The
parameter $\lambda$ is the bare screening length. A similar Hamiltonian
occurs for ceramic (granular) superconductors including the
self-inductance of vortex loops \cite{balseiro,kawamura,remark}.
For simplicity we set $J=1$.

After standard manipulations \cite{kleinert,natter} this Hamiltonian
can be brought into the vortex representation \cite{BY}
\be
H_{V}
= -\frac{1}{2}\sum_{i,j} ({\bf J}_i - {\bf b}_i) G(i-j) 
({\bf J}_j - {\bf b}_j)\;,
\label{vortex}
\ee
The ${\bf J}_i$, representing the vortex density of the phase fields
$\phi_i$ on bond $i$, are three-component integer variables running from
$-\infty$ to $\infty$, live on the links of the dual lattice and
satisfy the divergence constraint $({\bf \nabla} \cdot {\bf J})_i = 0$
on every site $i$. The function G(r) is a lattice Greens function
behaving asymptotically like $G(r)\sim r^{-1}\exp(-r/\lambda)$. The
${\bf b_i}$ are magnetic fields which are constructed from the
quenched vector potentials $A_{ij}$ by a lattice curl, in the present
case with the homogeneous external field we have:
\be
{\bf b}_i = \frac{1}{2 \pi}[\nabla \times {\bf A}^{\rm rand}]_i 
+ {\bf B}^{\rm ext}_i\;.
\label{b}
\ee
Obviously the random part fulfills the divergence free condition
$({\bf \nabla} \cdot {\bf b})_i = 0$. We specify the boundary
conditions for the vortex glass Hamiltonian to be periodic in all
space directions (corresponding to fluctuating boundary conditions in
the phase variables of the original gauge glass Hamiltonian
\cite{bc,gingras}.  Now we choose ${\bf B}^{\rm ext}_i=B\,{\bf e}_z$
i.e.\ the external field points in the $z$-direction (i.e.\ along
${\bf e}_z$ the unit vector in the $z$-direction) and is also
divergence-less due to the periodic boundary conditions.

We just remark that in the {\it pure} case (${\bf A}^{\rm rand}=0$)
the field strength $B$ simply plays the role of the usual filling
factor $f$ counting the number of flux units per plaquette giving rise
to the uniformly frustrated XY-model (see e.g.\ \cite{gingras,miya} and
references therein) in the unscreened case ($\lambda=\infty$). Here,
due to the long range interaction $G(r)\approx 1/r$, the ground state
is indeed non-trivial for irrational filling factors. In the continuum
limit the flux lines would actually form a hexagonal lattice, the well
known Abrikosov flux line lattice.

For the disordered case one has an interesting interplay between two
sorts of frustration: one is also present in the pure case and coming
from the external field and the other comes from the quenched
disorder. To our knowledge this problem has not been investigated
systematically so far. Here, as a first step, we confine ourselves to
the strongly screened case ($\lambda\to0$), for which the vortex
Hamiltonian (\ref{vortex}) simplifies \cite{WY} to
\be
H_V^{\lambda \rightarrow 0}=\frac{1}{2}\sum_i ({\bf J}_i - {\bf b}_i)^2\;.
\label{screen}
\ee
The problem of finding the ground state, i.e.\ the minimum energy
configuration of this Hamiltonian, is actually a
minimum-cost-flow-problem $\min_{\{{\bf J}_i\}} \sum_i c_i({\bf J}_i)$
subject to the constraint $(\nabla \cdot {\bf J})_i=0$, where
\mbox{$c_i({\bf J}_i) := ({\bf J}_i - {\bf b}_i)^2 /2$} are the so
called (convex) cost functions. This problem can be solved exactly
in polynomial time via combinatorial optimization techniques, as
described in \cite{review,KR}.

\section{Field chaos}
First we investigate how the ground state of the vortex glass changes
with the external field. Obviously, with increasing field strength $B$
more and more flux will flow in the $z$-direction. In Fig. 1 we depict
various quantities reflecting this observation: a) the total flux
$L^{-3}\sum_i (\vert J_i^x\vert + \vert J_i^y\vert + \vert J_i^z\vert)$,
which increases, after a crossover region, linearly with $B$; b) the
fraction of flow variables in the x- and y-direction that are
zero/nonzero; then the fraction of flow variable in the $z$-direction
which are c) larger than, d) smaller than, e) equal to zero.

\begin{figure}
\psfig{file=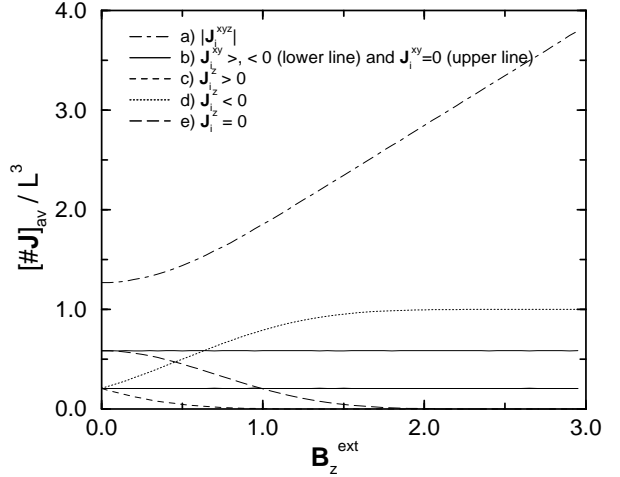,width=\columnwidth}
\caption{\label{Bc}
  Plot of the count of components per site vs. external field for 1000
  samples. a) The total flux $L^{-3}\sum_i (\vert J_i^x\vert + \vert
  J_i^y\vert + \vert J_i^z\vert)$; b) $L^{-3}\sum_i \delta_{J_i^{x},0}$ 
  (upper line), $L^{-3}\sum_i\theta (+J_i^{x})$ and 
  $L^{-3}\sum_i\theta (-J_i^{x})$ (both lay on the lower line); 
  the $y$-values are the same;  
  c) $L^{-3}\sum_i\theta (+J_i^z)$; d) $L^{-3}\sum_i\theta (-J_i^z)$;
  e) $L^{-3}\sum_i\delta_{J_i^z,0}$.  }
\end{figure}

For $B=0$ the vortex variables $J_i$ are 
homogeneously distributed; $L^{-3}\sum_i \delta_{J_i,0}$, 
$L^{-3}\sum_i\theta (+J_i)$ and 
$L^{-3}\sum_i\theta (-J_i)$ have the same values for each direction,
where the last two values are equal due to the symmetry of direction.
An increasing field $B>0$ only effects the components along the 
$z$-direction (dashed and dotted lines) and let the components 
perpendicular to it unchanged (solid lines).
There is a critical external field $B_c$, above which all
components of the flux lines ${\bf J}_i^{z}$ parallel to the field
have a non-vanishing negative amplitude, what is attended by a
decreasing of the zero-components, i.e. for $B\rightarrow B_c$ the
$\#\{ {\bf J}_i^{z} < 0\}$ increases (Fig. \ref{Bc}d) and the $\#\{
{\bf J}_i^{z} = 0\}$ decreases (Fig. \ref{Bc}c).  Before all
zero-components vanish ($B_c = 5/2$), the positive components have
disappeared ($B_c = 3/2$).  The critical fields can be determined by
remembering that the random field ${\bf b}^{ran}$ is in $[-2,+2]$ and
the flux is an integer.  Thus the maximal discrepancy between the
optimal flux and field is $1/2$ per site.  The optimal ${\bf J}_i^{z}$
depends on the applied field by a step function and in the worst case
the random field is ${\bf b}_i^{ran}= \pm 2$, so that we have $\{{\bf
  J}^{opt}\} = \{{\bf J}_i \quad \mbox{integer}| \quad \min|{\bf J}_i
- ({\bf B}_i^{ext} \pm 2.0 )| \quad\}$.  One obtains: ${\bf J}_i^{z} <
0$ $\forall i$ if $B>B_{c_1} = 5/2$ and ${\bf J}_i^{z} \le 0$ $\forall i$
if $B>B_{c_2} = 3/2$, in accordance with Fig. 1.

Next we study the sensibility of the ground state (or optimal flow
configuration) with respect to small changes in the external field
$B$. To this end we compare the ground state configurations of samples
with the same quenched disorder and slightly different external field
$B$. Denoting with ${\bf J_i}$ the zero field ($B=0$) ground state and
with ${\bf J_i}(B)$ the ground state of the same sample in
non-vanishing external field $B$ we define the Hamming distance of the
two configuration ${\bf J}_i$ and ${\bf J}_i(B)$ by
\begin{figure}
\psfig{file=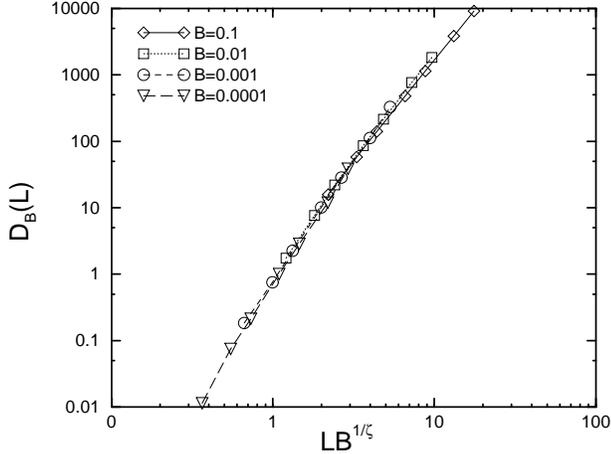,width=\columnwidth}
\caption{
\label{Chaos}
Scaling plot of the Hamming distance $D_B(L)$ vs.
$L B^{1/\zeta}$ for $L \le 32$: 5000 samples for $L\le 16$, 2000
for $L = 24$ and 500 for $L = 32$. The chosen values for $B$ are $B$ =
0.0001, 0.0010, 0.0100 and 0.1000. The best data collapse is achieved
by a chaos exponent $\zeta = 3.8 \pm 0.2$. The error bars are less 
than the size of the symbols and thus omitted.}
\end{figure}
\be D_B(L) = \sum_i({\bf J}_i(B) - {\bf J}_i)^2\;, 
\ee 
so that a small value of $D_B(L)$ means a strong correlation of
the ground states.

In \cite{KR} it has been found that an infinitesimal {\it random}
perturbation of the vector potential ${\bf A}_{ij}^{\rm rand}$ leads
to a chaotic rearrangement of the ground state configuration. There it
was demonstrated that, like in spin glasses\cite{Rie96,Ney97}, beyond
a particular length scale, the so-called overlap length, the two ground
states (perturbed and unperturbed) decorrelate. For the {\it
  non-random} magnetic field perturbation we study here we take over
this concept and demonstrate the existence of an overlap length $l^*$
scaling with the strength of the external field $B$ like $l^*\propto
B^{1/\zeta}$, where $\zeta$ is the chaos exponent.  For this length
scale to exist the finite size scaling form $D_B(L) =
d(L/l^*)=d(L/B^{1/\zeta})$ should hold, which is indeed satisfied, as
is shown in Fig. 2. We obtain a relatively large chaos exponent $\zeta
= 3.8 \pm 0.2$. Remarkably this exponent coincides (within the error
bars) with the chaos exponent for a {\it random} perturbation which
has been reported to be $\zeta^{\rm rand} = 3.9 \pm 0.2$ \cite{KR}.

\section{Defect energy (DWRG)}

In this section we study the scaling behavior of low energy
excitations $\Delta E(L)$ of length scale $L$ (to be defined below) in
the presence of an external field, which provides the essential
evidence about the stability of the ground state with respect to
thermal fluctuations. If $\Delta E(L)$ decreases with increasing
length $L$ it implies that it costs less energy to turn over larger
domains thus indicating the absence of a true ordered (glass) state at
any non-vanishing temperature. Usually one studies such excitation of
length scale $L$ by manipulating the boundary conditions (b.c.) for
the phase variables of the original Hamiltonian (\ref{gauge}), see
\cite{BY,kost2}. One induces a so called domain wall of length scale
$L$ into the system by changing the b.c. of a particular sample from
periodic to anti-periodic (or vice versa) in one space direction and
measures the energy of such an excitation by comparing the energy of
these two ground state configurations. This is the common procedure
for a DWRG analysis, which, however, bears some technical
complications \cite{BY} and some conceptual ambiguities
\cite{kost1,kost2} in it.

Here we follow the basic idea of DWRG, we will, however, avoid the
complications and the ambiguities that appear by manipulating the b.c.
and try to induce the low energy excitation in a different way, as it
has first been done by one of us in \cite{KR} for the zero-field case.

First we clarify what a low energy excitation of length scale $L$ is:
in the model under consideration here it is certainly a global vortex
loop encircling the 3d torus (i.e.\ the $L\times L\times L$ lattice
with periodic b.c.)  once (or several times) with minimum energy cost.
For the pure case the global minimum energy loop is simply a straight
line that costs energy $\Delta E(L)=JL$, which is exactly also what
one would expect for a domain wall of length $L$ in a three-dimensional
XY-model, and which is also obtained from the energy difference
between ground states with periodic and anti-periodic b.c.

Next we have to clarify how we induce the above mentioned global
vortex loop, if not by manipulating the b.c. The solution is to
manipulate instead the costs for flow in one particular space
direction. Suppose we have the exact ground state of one particular
sample, which specifies also the current cost for increasing the the
flow variables $J_i^{x,y,z}$ with respect to ${\bf J}^0$ by one
unit, e.g.: $\Delta c_i^x=c_i(J_i^{0x}+1)-c_i(J_i^{0x})
=J_i^{0x}-b_i^x+1/2$. If we decrease smoothly these variables $\Delta
c_i^x$ and apply our min-cost-flow algorithm to this modified problem
at some point a configuration ${\bf J}_i^1$ that is the original
ground state {\it plus} a global loop in the $x$-direction appear as
the new optimal flow configuration for the modified problem. This
extra loop, which can be easily identified by comparing the new
optimum with the original ground state, is the low energy excitation
we are looking for. Its energy $\Delta E(L)$ is simply the difference
of this state (ground state with loop), $H({\bf J}^1)$ minus the
energy of the ground state $H({\bf J}^0)$. Note that this is always
positive, since it is definitely an excitation (in contrast to the
usual DWRG procedure where the b.c.\ is modified).

\begin{figure}
\psfig{file=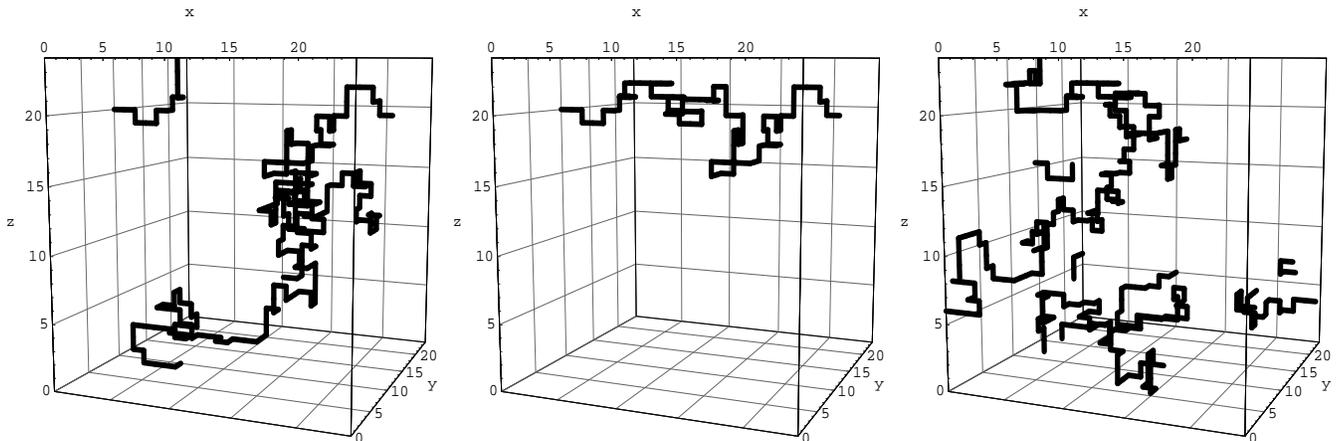,width=\columnwidth}
\caption{\label{dEB_24} The minimum energy global exciation loop {\it
  perpendicular} to the external field in the $z$-direction is shown
  for one particular sample ($L=24$) and three different field
  strengths $B$ (note the periodic b.c.\ in all space directions).
  {\bf a)} (left) $B\in [0.0065 , 0.0069]$ is in a range, where the
  defect energy $\Delta E$ varies linear with respect to the field
  (see inset of Fig. \ref{dEB_24gsmt}). Note that the loop has also
  winding number $n^z=1$ in the direction {\it parallel} to the
  external field. Hence $\partial \Delta E/\partial B = 2 L$. {\bf b)}
  (middle) The same sample as in (a) with $B\in [0.0070, 0.0075]$.  In
  this interval the defect energy is constant, no loop along
  the direction of the applied field occurs.  {\bf c)} (right) The
  same sample as in (a,b) with $B \in [0.0076 , 0.0081]$.  The system
  is very sensible to the variation of applied field $\Delta B$.  Even
  for a small change by $\Delta B = 0.0001$ the form of the excitation
  loop changes drastically (compare with a and b).}
\end{figure}

Four remarks are in order: 1) small, topologically simply connected
loops are not generated by this procedure, since all what can be
gained in energy is lost again on the return. 2) In the pure case this
procedure would not work, since at some point spontaneously {\it all}
links in the $z$ direction would increase their flow value by one. It
is only for the disordered case with a continuous distribution for the
random variables ${\bf b}_i$ that a unique loop can be expected. 3)
Sometimes (in ca.\ $5\%$ of the samples) the global flux changes
discontinuously by more than one unit, a typical example for such an
elementary excitation loop is shown in Fig. \ref{dEB_24}c, however we
define these still to be elementary excitations of length scale $L$.
4) In the presence of a homogeneous external field one has to
discriminate between different excitation loops: those parallel and
those perpendicular to the external field need not to have the same
energy (however, it turns out that the disorder averaged defect energy
is identical in all directions, see below).

Schematically the numerical procedure is the following:
\begin{itemize}
\item[1.] Calculate the exact ground state configuration \{${\bf
    J}^0$\} of the vortex Hamiltonian(\ref{vortex});
\item[2.] Determine the resulting global flux along, say, the $x$-axis
  $f_x = \frac{1}{L}\sum_i J_i^{0x}$;
\item[3.] Study a minimum-cost-flow problem in the actual cost for
  increasing the flow in the $x$-direction $\Delta
  c_i^x=c_i(J_i^{0x}+1)-c_i(J_i^{0x})-b_i^x+1/2$ is smmothly modified
  letting the cost of a topologically simple connected loop unchanged
  and only affecting gloabal loops.
\item[4.] Reduce the $\Delta c_i^x$ until the optimal flow
  configuration \{${\bf J}^1$\} for this min-cost-flow problem has the
  global flux $(f_x + 1)$, corresponding to the so called $elementary$
  low energy excitation on the length scale $L$;
\item[5.] Finally, the defect energy is $\Delta E=H(\{{\bf
    J}^1\})-H(\{{\bf J}^0\})$.
\end{itemize}

It turns out that the computation time grows linearly with $B$ and it
took about 150 seconds per samples for $L=24$ with $B=1$ on a SUN
workstation (167MHz). \\

As for the zero-field case \cite{KR} one expects for the disordere
averaged excitation energy (or defect energy)
\be
[\Delta E(B,L)]_{\rm av} \sim L^{\theta}\;,
\label{scaling}
\ee
where $B$ is fixed, $[\cdots]_{\rm av}$ denotes the disorder average
and $\theta$ is the stiffness exponent and its sign determines whether
there is a finite temperature phase transition or not, as explained
above. If $\theta<0$, i.e.\ the transition to a true superconducting
vortex state appears only at zero temperature $T=0$, scaling predicts
that the thermal correlation length $\xi$ diverges at $T=0$ as
$\xi\propto T^{-\nu}$ with $\nu=1/\vert\theta\vert$, c.f.\ 
\cite{WY,KR,Rie96}.

In the non-random case ${\bf A}_{ij}^{\rm rand}=0$ an elementary
exciation of the kind we described, i.e.\ a global loop of length $L$
around the torus costs an energy of $\Delta E \sim L^{+1}$ without any
dependence of the external field as well parallel as well as
perpendicular to the external field (i.e.\ the $z$-direction).  Note,
that this is the situation for of the pure classical 3d XY-model with
strong screening.

\begin{figure}
\psfig{file=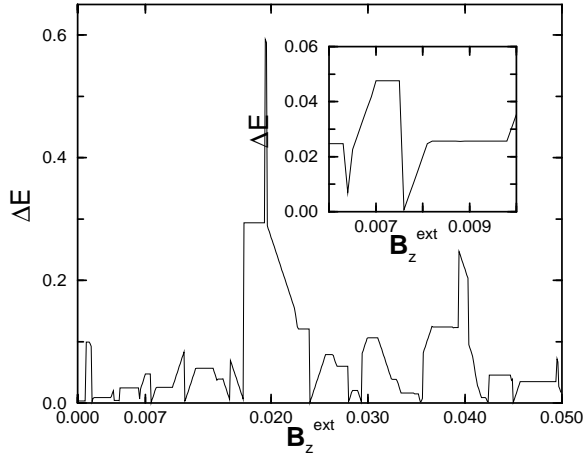,width=\columnwidth}
\caption{
   \label{dEB_24gsmt}Defect energy $\Delta E$ vs. applied magnetic field 
   $B$ for one particular disorder configuration.  The field varies
   between $0$ and $0.05$ times one flux unit and the system size is
   $L=24$.  The inset enlarges the region that is studied in Fig. \ref{dEB_24}
   in more detail.
}
\end{figure}

For a single configuration we find that a change of the external magnetic
field $B$ drastically affects the defect energy $\Delta E$
(Fig. \ref{dEB_24gsmt}). $\Delta E$ is a piecewise linear function that
behaves in particular intervals $[B_a,B_b]$ as
\be
  \frac{\partial \Delta E}{\partial B} = 2 L n^z\;,
\ee
\end{multicols}
\widetext
\noindent\rule{20.5pc}{.1mm}\rule{.1mm}{2mm}\hfill
\begin{multicols}{2} 
\narrowtext
\noindent 
\noindent
which can be understood as follows: the external field varies
continuously and the interger valued flow changes 
only in discrete steps, thus the minimum energy exciation loop may not
change in a whole interval say $[B_a,B_b]$. In this interval $\Delta
E$ changes linearly with $B$ since $H({\bf J}^1)-H({\bf J}^0)$ is
simply proportianl to length of the exciation loop in the
$z$-direction, which is $n^z\cdot L$, with $n^z$ the winding number of
the loop in the $z$-direction
($n^z\in\{\ldots,-2,-1,0,+1,+2,\ldots\}$).

Furthermore not only the ground state itself is extremely sensible to
the small variations of the external field strength (as we have seen
further above), but also the excitation loops themselves change their
form dramatically, as it is exemplified in Fig. \ref{dEB_24}. Only
small parts of the loop seem to persist over a significant range of
the field strength, see for instance the in the vicinity of the plane
$z=20$ in Fig. \ref{dEB_24}.

\begin{figure}
\psfig{file=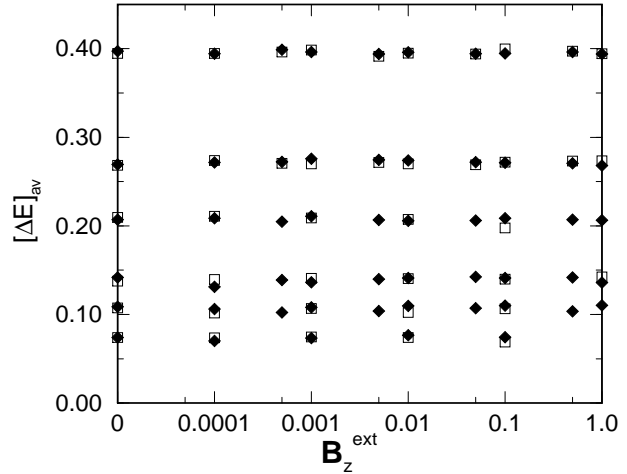,width=\columnwidth}
\caption{\label{45} Linear plot of the external magnetic field vs. 
  the defect energy [$\Delta E$]$_{av}$ for $L = 4,6,8,12,16,24$ (top
  to button).  The elementary excitation parallel $\Box$ and
  perpendicular full diamond to the external magnetic field
  ${\bf B}^{ext}_z$.  For each plotted point the number of samples
  varied between 20000 for the smallest sizes to 1000 for the largest
  sizes.  The error bars of the excitations are less than the size of
  the symbols and thus omitted.  }
\end{figure}

Now we study the behavior of the disorder averaged energy of excitation
loops perpendicular to the applied magnetic field along, say, the
$z$-direction (full diamonds in Fig. \ref{45}). Note that it is only
necessary to study the situation $B\in[0,1]$, since all physical
properties of the vortex glass Hamiltonian (\ref{screen}) are periodic
in the strength of the external field $B$, i.e.\ the filling factor.
As can easily be seen in Fig. \ref{45} the defect energy $[\Delta
E(B,L)]_{\rm av}$ is independent of the value of $B$. For any fixed
value of $B$ the finite size scaling relation (\ref{scaling}) is
confirmed and gives $\theta= -0.95 \pm 0.04$, c.f.\cite{KR}.

We want to note that this behavior of excitations perpendicular to the
applied field depends neither on the length of the system in the
$z$-direction nor on the topology in this direction: we also studied
the situation in which the vortex Hamiltonian (\ref{screen}) lives on a
lattice with free instead of periodic b.c.\ in the $z$-direction. In
this
case the external field has an appropriate source and sink
outside the system. We find here the same result as before: the
disorder averaged defect energy is independent of the strength of the
external field $B$.

Finally, we also observe a field-independent domain wall energy for
elementary excitation loops (open boxes in Fig. \ref{45}) {\it
  parallel} to the external field. Thus we conclude that the disorder
averaged defect energy is also independent of the direction of the
homogeneous external field.

\section{Summary}
We have studied the three-dimensional vortex glass model in the strong
screening limit in the presence of a homogeneous external field. The
ground state is extremely sensible to small external field
variations. Ground state configurations at different field values $B$
and $B+\Delta B$ decorrelate beyond the overlap correlation length
$l^*\sim\Delta B^{1\zeta}$, where $\zeta$ is the so called chaos
exponent which we estimate to be $\zeta=3.8\pm0.2$. This value agrees
within the error-bars with the chaos exponent for {\rm random}
perturbations of the quenched disorder that have been investigated in
\cite{KR}.

For individual disorder configurations the change of the defect energy
$\Delta E_L(B)$ with respect to the applied field $B$ is piecewise
linear, analytic and accompanied by a drastic deformation of the
minimum energy global excitation loop. On the other hand the disorder
averaged value of the defect energy $[\Delta E_L(B)]_{\rm av}$ is
independent of the strength of the external field $B$. Moreover it
turned out that also the excitation loops parallel as well as
perpendicular to the external field yield the same disorder averaged
value $[\Delta E_L(B)]_{\rm av}$. Thus the scaling behavior of the
defect energy is independent of $B$, i.e.\ iidentical to the case $B=0$
already studied in \cite{KR}. Therefore, as in the $B=0$ case, we
infer the absence of a true superconducting low temperature phase in
the strongly screened vortex glass model in an external field. The
stiffness exponent is again, as in the $B=0$ case,
$\theta=-0.95\pm0.04$, giving an estimate
$\nu=1/\vert\theta\vert=1.05\pm0.05$ for the thermal correlation
length exponent.

Concluding we would like to note that it would be interesting to
perform the same analysis for non-vanishing screening length and for
the unscreened case, where due to the long range repulsion of the
vortex lines important new physics might appear. In particular it is
an open whether a homogeneous external field has a significant effect
on the existence of the low temperature vortex glass phase in the
unscreened case.

{\bf Acknowledgment} We thank J. Kisker for important help during
this work.  HR's work has been supported by the Deutsche
Forschungsgemeinschaft (DFG).

\end{multicols}


\begin{references}

\bibitem{blatter}
  G.\ Blatter et al., Rev. Mod. Phys. {\bf 66}, 1125 (1994).

\bibitem{Fisher}
  M.~P.~A.~Fisher, Phys. Rev. Lett. {\bf 62}, 1415 (1989).

\bibitem{FFH}
  D.~S.~Fisher, M.~P.~A.~Fisher and D.~A.~Huse, 
  Phys. Rev. B {\bf 43}, 130 (1991).

\bibitem{exp}
  R. H. Koch et al.,
  Phys. Rev. Lett. {\bf 63}, 1512 (1989);
  P. L. Gammel, L. F. Schneemeyer and D. J. Bishop,
  Phys. Rev. Lett. {\bf 66}, 953 (1991);
  T. K. Worthington et al.,
  Phys. Rev. B {\bf 43}, 10538 (1991);
  C. Dekker, W. Eidelloth, and R. H. Koch,
  Phys. Rev. Lett. {\bf 68}, 3347 (1992);
  Y. Ando, H. Kubota, and S. Tanaka,
  Phys. Rev. Lett. {\bf 69}, 2851 (1992).
  
\bibitem{FTY}
  M.~P.~A. Fisher, T.~A. Tokuyasu and A.~P. Young, 
  Phys. Rev. Lett. {\bf 66}, 2931 (1991).

\bibitem{kost1}
  J. M. Kosterlitz and M. V. Simkin,
  Phys. Rev. Lett. {\bf 79}, 1098 (1997).

\bibitem{kost2}
  J. M. Kosterlitz and N. Akino,
  Phys. Rev. Lett. {\bf 81}, 4672 (1998).

\bibitem{BY}
  H.~S.~Bokil and A.~P.~Young, 
  Phys. Rev. Lett. {\bf 74}, 3021 (1995).

\bibitem{WY}
  C.~Wengel and A.~P. Young, 
  Phys. Rev. B {\bf 54}, R6869 (1996).

\bibitem{KR}
  J. Kisker and H. Rieger,
  Phys. Rev. B {\bf 58}, R8873 (1998).

\bibitem{gauge}
  C.\ Ebner and D.\ Stroud,
  Phys. Rev. B {\bf 31}, 165 (1985).
  S.\ John and T.\ C.\ Lubensky,
  Phys. Rev. B {\bf 34}, 4815 (1986).

\bibitem{gremp}
  J. Maucourt and D. R. Grempel,
  Phys. Rev. Lett. {\bf 80}, 770 (1998).

\bibitem{balseiro}
  D. Dominguez, E. A. Jagla, and C. A. Balseiro,
  Phys. Rev. Lett. {\bf 72}, 2773 (1994).

\bibitem{kawamura}
  H. Kawamura and M. S. Li,
  Phys. Rev. B {\bf 54}, 619 (1996);
  J. Phys. Soc. Jap. {\bf 66}, 2110 (1997).

\bibitem{remark} Note however that in
  \protect{\cite{balseiro,kawamura}} the disorder is modelled by
  random Josepson couplings (here constant $J$), not by random vector
  potentials. Therefore the model considered there is a XY spin glass
  model rather than a gauge glass model.

\bibitem{kleinert} 
  J.~Villain, J. Physique {\bf 36}, 581 (1975);
  J.~V.~ Jos\'e, L.~P.~Kadanoff, S.~K.~Kirkpatrick, and D.~R. Nelson,
  Phys. Rev. B {\bf 16}, 1217 (1977);
  H.~Kleinert, {\em Gauge fields in Condensed Matter}, (World Scientific, 
  Singapore, 1989).

\bibitem{natter} 
  M.\ S.\ Li,  T.\ Nattermann, H.\ Rieger und M.\ Schwartz,
  Phys.\ Rev.\ B {\bf 54}, 16024 (1996).

\bibitem{bc}
  P.~Olsson, Phys. Rev. B {\bf 52}, 4511 (1995);

\bibitem{gingras}
  P.~G.~ Gupta, S.~Teitel and M.~J.~P.~Gingras,
  Phys. Rev. Lett. {\bf 80}, 105 (1998).

\bibitem{teitel}
  M. Franz and S. Teitel,
  Phys. Rev. Lett. {\bf 73}, 480 (1991)

\bibitem{miya}
  X. Hu, S. Miyashita, and M. Tachiki,
  Phys. Rev. Lett. {\bf 79}, 3498 (1997).

\bibitem{review}
  H.~Rieger, {\em Frustrated Systems: Ground State Properties via
  Combinatorial Optimization}, Lecture Notes in Physics 501 
  (Springer Verlag Heidelberg, 1998);
  M.\ Alava, P.\ Duxbury, C.\ Moukarzel, H.\ Rieger,
  {\it Combinatorial optimization and disordered systems},
  to appear in ``Phase Transition and Critical Phenomena''
  ed.\ C. Domb and J. L. Lebowitz (Academic Press, 1999).
 
\bibitem{Rie96}
  U.~Blasum M.~Diehl H.~Rieger, L.~Santen and M.~J\"unger.
  J. Phys. A, {\bf 29}, 3939 (1996).

\bibitem{Ney97}
  M.~Ney-Nifle and A.~P. Young.
  J. Phys. A, {\bf 30}, 5311 (1997).

\end{references}
\end{document}